\documentclass[sigconf]{acmart}

\usepackage{tikz}
\usepackage{pgfplots}
\usepackage{subcaption}
\usepackage{booktabs}
\usepackage{algorithm}
\usepackage{algpseudocode}
\usepackage{seqsplit}
\usepackage{listings,multicol}
\usepackage[templates]{genealogytree}
\usepackage{balance}
\usepackage{paralist, enumitem}

{}
{}
{}

\def\HiLi{\leavevmode\rlap{\hbox to \hsize{\color{yellow!50}\leaders\hrule height .8\baselineskip depth .5ex\hfill}}}
\definecolor{codegreen}{rgb}{0,0.5,0}
\begin{document}

%\supertitle{Submission Template for IET Research Journal Papers}

\title{A unit-based symbolic execution method for detecting memory corruption vulnerabilities in executable codes}

\author{Sara Baradaran}
\affiliation{%
 \institution{Isfahan University of Technology, }
 \city{Isfahan}
 %\state{Arunachal Pradesh}
 \country{Iran}
 }%\email{s.baradaran@ec.iut.ac.ir}

\author{Mahdi Heidari}
\affiliation{%
 \institution{Isfahan University of Technology, }
 \city{Isfahan}
 %\state{Arunachal Pradesh}
 \country{Iran}}

\author{Ali Kamali}
\affiliation{%
 \institution{Isfahan University of Technology, }
 \city{Isfahan}
 %\state{Arunachal Pradesh}
 \country{Iran}}

\author{Maryam Mouzarani}
\affiliation{%
 \institution{Isfahan University of Technology, }
 \city{Isfahan}
 %\state{Arunachal Pradesh}
 \country{Iran}}

\renewcommand{\algorithmicrequire}{\textbf{Input:}}
\renewcommand{\algorithmicensure}{\textbf{Output:}}

\settopmatter{printacmref=false}
%\author{\au{Sara Baradaran \orcidlink{0000-0002-9148-6493}$^{1}$}, \au{Mahdi Heidari \orcidlink{0000-0003-4342-6637}$^{1}$}, \au{Ali Kamali \orcidlink{0000-0002-3913-1329}$^{1}$}, \au{Maryam Mouzarani \orcidlink{0000-0002-8134-3042}$^{1\corr}$}}

%\author{\au{Sara Baradaran$^{1}$}, \au{Mahdi Heidari$^{1}$}, \au{Ali Kamali$^{1}$}, \au{Maryam Mouzarani$^{1\corr}$}}

%\address{\add{1}{Department of Electrical and Computer Engineering, Isfahan University of Technology, Isfahan, Iran} 

%\email{s.baradaran@ec.iut.ac.ir, heidari@ec.iut.ac.ir, a.kamali@ec.iut.ac.ir, mouzarani@iut.ac.ir}}

\begin{abstract}

Memory corruption is a serious class of software vulnerabilities, which requires careful attention to be detected and removed from applications before getting exploited and harming the system users. Symbolic execution is a well-known method for analyzing programs and detecting various vulnerabilities, e.g., memory corruption. Although this method is sound and complete in theory, it faces some challenges, such as path explosion, when applied to real-world complex programs. In this paper, we present a method for improving the efficiency of symbolic execution and detecting four classes of memory corruption vulnerabilities in executable codes, i.e., heap-based buffer overflow, stack-based buffer overflow, use-after-free, and double-free. We perform symbolic execution only on test units rather than the whole program to lower the chance of path explosion. In our method, test units are considered parts of the program's code, which might contain vulnerable statements and are statically identified based on the specifications of memory corruption vulnerabilities. Then, each test unit is symbolically executed to calculate path and vulnerability constraints of each statement of the unit, which determine the conditions on unit input data for executing that statement or activating vulnerabilities in it, respectively. Solving these constraints gives us input values for the test unit, which execute the desired statements and reveal vulnerabilities in them. Finally, we use machine learning to approximate the correlation between system and unit input data. Thereby, we generate system inputs that enter the program, reach vulnerable instructions in the desired test unit, and reveal vulnerabilities in them. This method is implemented as a plugin for \textit{angr} framework and evaluated using a group of benchmark programs. The experiments show its superiority over similar tools in accuracy and performance. 
\end{abstract}

\maketitle

\textbf{Accepted Manuscript} - 	International Journal of Information Security - https://doi.org/10.1007/s10207-023-00691-1

\section{Introduction}\label{sec1}

Memory corruption vulnerabilities are prevalent and detrimental software weaknesses, which potentially occur when programming with low-level languages (usually C and C++). These vulnerabilities allow attackers to access the program's memory directly and read from or write to it arbitrary values. Although low-level languages provide few security guarantees \cite{refbib4}, their flexibility and efficiency encourage programmers to use them in developing critical software, such as operating system kernels, drivers, and OS services. Therefore, exploiting memory corruption vulnerabilities might seriously harm the system owners.

A wide variety of program analysis and vulnerability detection methods have been introduced over the past decades. Among them, symbolic execution is a promising one, which systematically explores the program execution paths with high coverage. In this method, input values are represented as symbols to calculate symbolic constraints on input data for each execution path and generate sufficiently diverse test data for analyzing a program's possible behaviors \cite{refbib1}. Although symbolic execution is theoretically sound and complete \cite{refbib2}, it may run into challenges in analyzing real-world programs. A well-known example of these challenges is path explosion that emerges since the number of program execution paths grows exponentially, and it makes storing and exploring the paths of large programs infeasible.

Some researchers have applied machine learning techniques to improve symbolic execution and prevent path explosion ~\cite{refbib7,refbib11,refbib6,refbib5,refbib19,refbib3}. For instance, in \cite{refbib19}, symbolic execution is applied to a test unit rather than the entire program to limit the scope of symbolic analysis and avoid path explosion. In the suggested method, a combination of concrete and symbolic execution is applied to calculate the constraints of execution paths in each test unit and generate appropriate test data to explore these paths by solving the calculated constraints. Then, curve fitting technique \cite{refbib10} is employed to approximate the correlation between system inputs and the test unit inputs as a function. Using this function, new system input data are generated to reach the test units and traverse various execution paths in them. This method is not used to detect a specific class of vulnerability and it does not contain details on how to determine the test units in a program.

We extend the idea presented in \cite{refbib19} and our recent work \cite{our} to propose a method for detecting four classes of memory corruption vulnerabilities in executable codes, i.e., heap-based buffer overflow, stack-based buffer overflow, use-after-free, and double-free. We present formal specifications for these vulnerability classes to be used to automatically locate test units in executable codes. Given the specification of a vulnerability class, test units are symbolically executed to calculate path and vulnerability constraints of the desired execution paths in each test unit. Then, a set of unit input data is generated by solving the calculated constraints to explore test units, reach vulnerable statements, and activate their vulnerabilities. Similar to the method in \cite{refbib19}, we estimate the relationship between the program and unit inputs by simulating the program execution and using machine learning techniques. Thereby, we generate test data that enter into the program from the beginning and reveal vulnerabilities in the desired instructions of a test unit.

The proposed method is implemented as a plugin for \textit{angr} \cite{refbib4}, which is a flexible, modular, and scalable binary analysis framework, supporting a variety of architectures. We have named our implemented method \textit{UbSym} as it employs \textbf{U}nit-\textbf{b}ased \textbf{Sym}bolic execution for detecting vulnerabilities in executable codes. The performance and accuracy of \textit{UbSym} have been evaluated using a group of NIST SARD benchmark vulnerable programs \cite{refbib30} and a number of complex programs that contain more functions and more complicated path constraints in comparison with the benchmark programs. We have also compared \textit{UbSym} with two tools, MACKE \cite{refbib24} and Driller \cite{refbib25}, which detect memory corruption vulnerabilities with similar methods. The experimental results show that our method performs more efficiently than these tools for detecting such vulnerabilities. The source code of \textit{UbSym}, including the test cases and scripts to automatically reproduce test results, is publicly available in our github repository \cite{refbib31}.

In summary, this paper makes the following contributions:

\begin{itemize}[leftmargin=*]
    \item {Specifying four vulnerability classes, i.e., stack-based buffer overflow, heap-based buffer overflow, double-free, and use-after-free, in executable codes and presenting a method to automatically determine test units for each vulnerability class in a program accordingly.}
    \item {Revising the algorithm presented in \cite{refbib19} to calculate path and vulnerability constraints based on our vulnerability specifications and focus on detecting memory corruption vulnerabilities more efficiently.}
    \item {Implementing and evaluating the total solution to demonstrate the effectiveness of unit-based symbolic execution against similar methods for detecting memory corruption vulnerabilities.}
\end{itemize}

The remainder of this paper is structured as follows: Section~\ref{related} discusses some related work. In Section~\ref{sec2}, the proposed method is described in detail. Section~\ref{sec3} explains our experiments and evaluates the implemented method. In Section~\ref{diss}, we mention some limitations of the proposed method, and finally, Section~\ref{sec4} concludes the paper.

\begin{figure*}[t]
\centering{\includegraphics[width=0.8\textwidth]{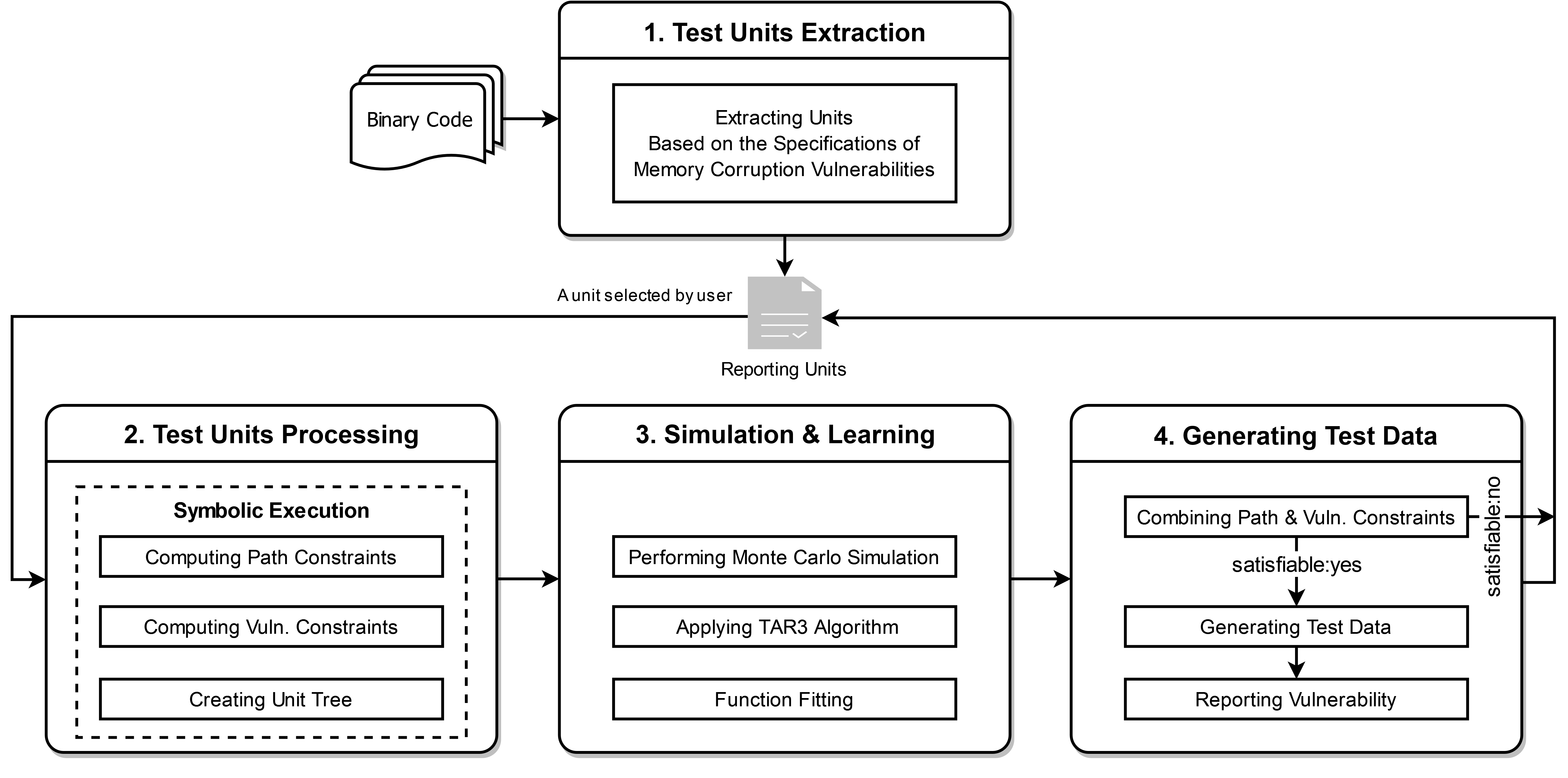}}
\caption{Architecture of the proposed method\label{fig:fig1}}
\end{figure*}

\begin{figure}[t]
\centering{\includegraphics[width=0.48\textwidth]{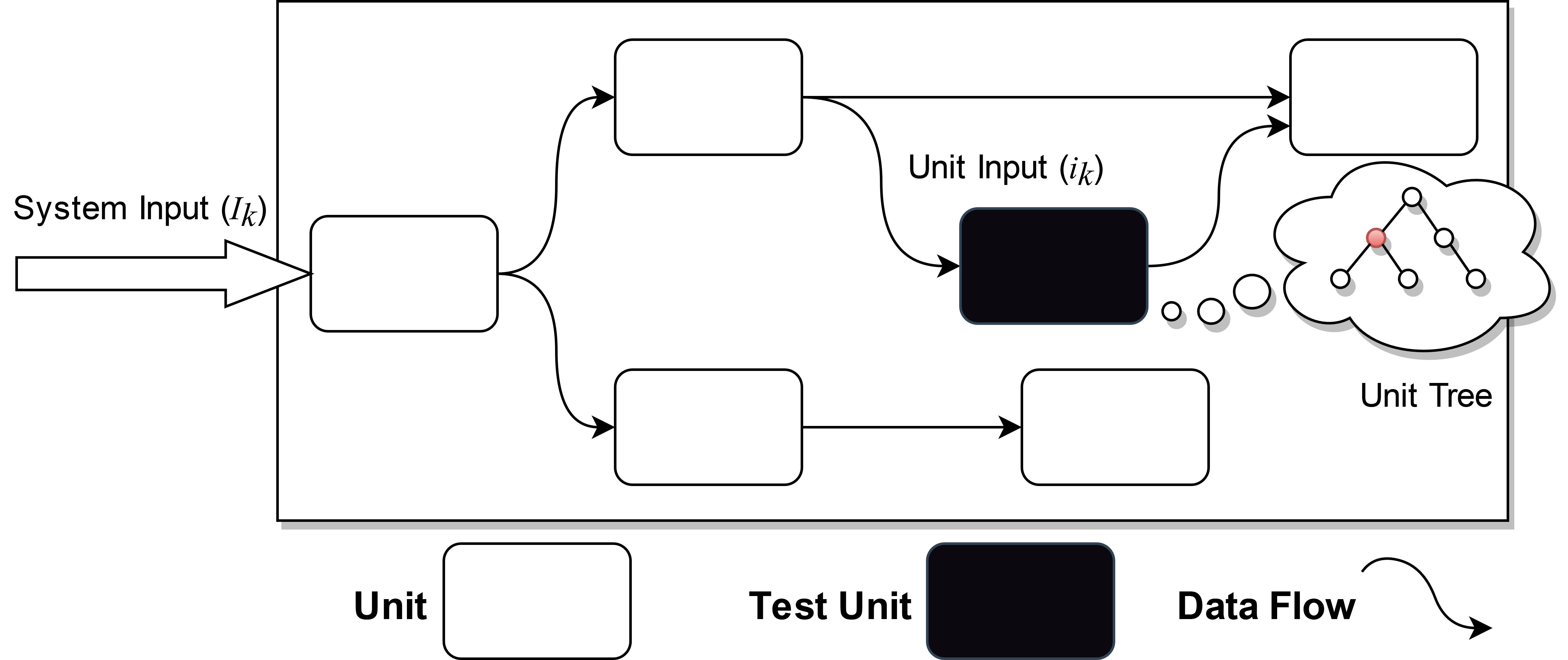}}
\caption{Schema of a program as a system containing a vulnerable unit with an input $i_k$ having a relevant system input $I_k$ obtained from curve fitting and treatment learning\label{fig:fig2}}
\end{figure}

\section{Related Work}
\label{related}
In recent years, various methods have been proposed by the scientific community to discover memory corruption vulnerabilities in programs \cite{Undangle, 13, 14, 15, refbib24, 19, 24, refbib25}. However, most of these methods have not been implemented to identify vulnerabilities in executable codes and only use source codes \cite{13, refbib24}. In many cases, executable code analysis is the only possible way to prove or disprove properties of the code that is actually executed \cite{refbib4}. In addition, some of these methods
are only used to discover specific vulnerability classes \cite{Undangle, 14, 15}. For example, the authors in \cite{Undangle} only focus on diagnosing use-after-free and double-free vulnerabilities. Similarly, the method presented in \cite{14} only targets heap-based buffer overflow vulnerability. When we come to use these methods for detecting other vulnerability classes, significant changes to the implemented algorithms are required. 

Our method discovers vulnerabilities in executable codes, which is much more accurate than analyzing the source code. Moreover, by providing the corresponding specification of a new vulnerability class, the presented method could be extended to explore test units for that vulnerability class and detect possible vulnerabilities in a program accordingly. In this section, we discuss some methods that have been provided by researchers for detecting memory corruption vulnerabilities. 

In \cite{13}, the authors proposed a comprehensive vulnerability detection approach for memory corruption vulnerabilities in programs written in C/C++. Their approach identifies unsafe operations, including both invalid memory writes and reads, in source code by static analysis based on safety constraints, involving the flow-sensitive point-to analysis and AST analysis on LLVM. The authors evaluated their approach using publicly available benchmarks and test suites. The experimental results showed an acceptable accuracy of detection. However, this paper does not contain details about the detection time, and the implemented approach is not open source.

The S2EDroid presented in \cite{24} uses selective symbolic execution for detecting memory corruption vulnerabilities in Android binary software. The authors first define the accession security rules for stack and heap memory. Then using selective symbolic execution, they check all the memory operations for all the execution paths to determine whether an illegal memory accession exists.

In \cite{refbib25}, the authors presented Driller as a vulnerability detection tool that combines dynamic fuzzing and concolic execution to efficiently discover vulnerabilities in binary codes. Driller uses evolutionary algorithms to generate multiple input values from an initial seed and explore the program paths. If the process is trapped in a part of the program because of a complex conditional statement and the fuzzer fails to generate consistent input values for that condition, selective symbolic execution is applied to calculate the constraint and generate appropriate input data detecting more in-depth vulnerabilities. Driller is also among the most popular vulnerability detection tools, given its satisfactory performance in detecting vulnerabilities. 

In \cite{refbib24}, MACKE is presented as a framework written on top of the KLEE symbolic execution engine \cite{refbib33} for analysis of C programs and detecting unhandled memory operations that result in memory out-of bounds errors. MACKE first recognizes each function through static analysis as a test unit. Then, it performs symbolic execution on individual units, in isolation. Since MACKE does not consider the constraints of the path from the beginning of the program to the test units, it might generate several false positives in this step. Therefore, it statically analyzes the call graph and the program’s control flow graph to identify feasible function call scenarios and omit the vulnerabilities that could not be exploited. 

In Section~\ref{sec3}, \textit{UbSym} is compared with Driller and MACKE since these open source tools employ similar techniques to improve the coverage of analysis. Driller applies selective symbolic execution and uses the angr framework when the fuzzer fails to go in depth. MACKE
also breaks the program into several units and separately applies symbolic execution to each unit in order to enhance the coverage of analysis.

\section{Method Overview}\label{sec2}

Our proposed method is illustrated in Fig.~\ref{fig:fig1}. It consists of four phases: test unit extraction, test unit processing, learning and simulation, and generating test data. 

In the first phase, the program's executable code is statically analyzed to identify test units based on the specification of each memory corruption vulnerability. To make the process clearer, Fig.~\ref{fig:fig2} illustrates a program containing various units, which \textit{UbSym} identifies its test units, shown in black, during the static analysis in the first step. To dynamically detect vulnerabilities in such a program, we are interested in finding an input data $i_k$ for a test unit and its relevant system input data $I_k$ that causes vulnerability activation in the suspicious statements of that unit. Thus, in the second phase, we analyze the test units with symbolic execution and consider the rest of the program a black box. In this step, we generate a constraint tree for each test unit, which contains path constraints on unit input data for each node and vulnerability constraints on unit input data for nodes suspected to have memory corruption vulnerability. In the third phase, we perform Monte Carlo simulation and execute the whole program with multiple system input values. If system input $I_k$ reaches the test unit with input value $i_k$ and causes the execution of node $n$ in the unit constraint tree, we annotate node $n$ with the pair $(I_k, i_k)$ to record which input data cause executing that node. Then, for each possibly vulnerable node in the unit tree whose constraints are satisfiable, we use function fitting technique \cite{refbib12} to estimate the relation between system and unit input data as a function. Finally, in the fourth phase, we use the calculated path and vulnerability constraints and the estimated function to generate appropriate system input data that enter the program, reach the test unit, and cause vulnerability activation in vulnerable statements. In the following, we explain each phase in more detail.

\begin{figure*}[t]
\centering{\includegraphics[width=0.85\textwidth]{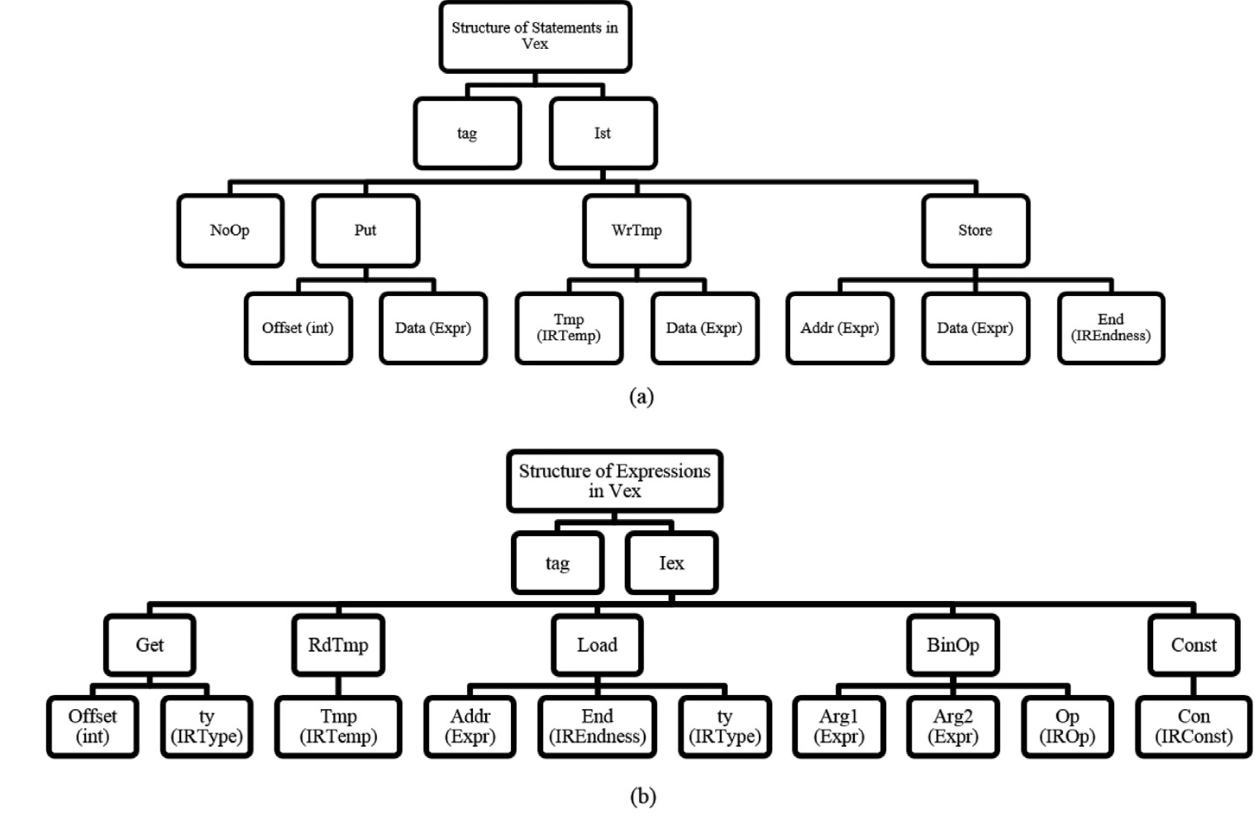}}
\caption{The structure of statements and expressions in VEX. Part (a) shows the structure of some of the statements in VEX. Part (b) shows the structure of some of the
expressions in VEX \cite{refbib21}.\label{fig:fig3}}
\end{figure*}

\subsection{Test units extraction}\label{subsec2.1}

In the first phase, the program's executable code is statically analyzed to search for functions that probably contain vulnerable statements. In order to locate possible vulnerabilities, we first specify how each vulnerability class appears in executable codes. We use the general vulnerability specification method presented in \cite{refbib21} to describe memory corruption vulnerabilities. In this method, a vulnerability is described as a single or a sequence of events. Each event is represented as a pair of \{CONT, Rule\}, which \{CONT\} defines the location of data in a program statement that is of concern in a vulnerability class and the \{Rule\} defines the conditions on CONT data, which result in intended vulnerability activation.

It is worth mentioning that since we implement our method in \textit{angr} framework and it translates binary instructions into VEX intermediate language, we define the containers and rules in our specifications based on the structure of VEX language. \textit{Angr} translates each assembly instruction into one or multiple statements in VEX language. VEX statements have different types, such as \texttt{Store}, \texttt{WrTmp}, etc. In addition, each statement might contain one or more expressions of different types, such as \texttt{Load}, \texttt{Get}, etc. Fig.~\ref{fig:fig3} presents the structure of the most common statements and expressions in VEX. Using the proposed method in \cite{refbib21} and based on VEX intermediate language, we specify four multi-event vulnerability classes, i.e., stack-based buffer overflow, heap-based buffer overflow, double-free, and use-after-free. In our specification, the symbol $\rhd$ represents the order of events.

\subsubsection{Test unit of heap-based buffer overflow vulnerability}\label{subsubsec2.1.1} 

Our proposed specification for heap-based buffer overflow vulnerability is presented in Fig.~\ref{fig:fig4}. This specification consists of two events: allocating a heap buffer and storing some data in that buffer.

\begin{figure}[t]
    \centering
    \scriptsize
    \fbox{\begin{minipage}{0.46\textwidth}
        \centering
        (\{\texttt{CONT1},\texttt{CONT2}\},True) $\rhd$ (\{\texttt{CONT3},\texttt{CONT4}\},Rule1)\\
        \vspace{5pt}
        \hrule
        \begin{enumerate}
            \item \texttt{CONT1 = malloc(CONT2)}
            \item \texttt{STle(CONT3) = CONT4\\
            \hspace{6pt} \textbf{\color{codegreen} Container[HOF].tag = Ist\_Store\\
            \hspace{6pt} Container[HOF].Store.addr = CONT3\\
            \hspace{6pt} Container[HOF].Store.data->tag = Iex\_RdTmp\\
            \hspace{6pt} Container[HOF].Store.data->Iex.RdTmp.tmp = CONT4\\
            }}
        \end{enumerate}
        \raggedright Rule1:\\
        \centering
        \texttt{CONT1 $\leq$ CONT3 $<$ CONT1 $+$ CONT2\\}
        \texttt{ $\wedge$ }
        \texttt{len(CONT4) > CONT1 + CONT2 - CONT3}
        \vspace{1pt}
        \end{minipage}}
    \caption{\centering Specification of heap-based buffer overflow vulnerability in VEX language\label{fig:fig4}}
\end{figure} 

In the first event of this vulnerability specification, we locate \texttt{malloc} function calls and consider their length arguments and return values as \texttt{CONT2} and \texttt{CONT1}, respectively. Here, \texttt{CONT1} represents the local variable storing address of the allocated buffer.

In the second event of the specification, we look for \texttt{Store} statements and check the source and destination of these statements. To clarify the specification, the detailed structure of \texttt{ Store} statement is represented in green in Fig.~\ref{fig:fig4}. As shown in this structure, the source and destination of the \texttt{Store} statement are defined in the \texttt{Store.data} and \texttt{Store.addr} sections, respectively. Also, the \texttt{Store.data} section is an expression with type \texttt{RdTmp}, in which the \texttt{RdTmp.tmp} part refers to the temporary variable containing the source data of the store operation and is considered as \texttt{CONT4}. Moreover, the destination buffer address, which is defined by \texttt{Store.addr} part of the \texttt{Store} statement, is considered as \texttt{CONT3}. Regarding to Rule1, heap-based buffer overflow occurs when a store operation is performed to a memory location in the range of an allocated heap buffer and length of the source data is more than size of the destination heap buffer, or \texttt{length(CONT4)} > \texttt{CONT2}. 

To determine the test units, we identify allocated heap buffers (\texttt{CONT1}) in executable codes and search for functions in which some data is stored in these buffers. Such functions are considered test units for this vulnerability. It is worth mentioning that since the address of a heap buffer is dynamically assigned at run time, we refer to the local pointer that stores the address of the allocated heap buffers in our static analysis. Since this pointer is a local variable and located in the stack memory, it is statically available and helps to follow the usage of heap buffers throughout the program statements and functions. We also assume that the length of the allocated heap buffer in the \texttt{malloc} function is a constant value and is available during the static analysis.

Another issue in static tracing of heap buffers arises in nested function calls. A heap buffer may be sent as an argument to other functions, and thus we need to trace it among the stacks of called functions. As a solution, for heap buffers passed to other functions or returned from function calls, we use the calling conventions in assembly languages to locate the local pointers holding the addresses of these buffers. The same challenge exists in identifying test units of the other memory corruption vulnerability classes, which is solved similarly.

\subsubsection{Test unit of stack-based buffer overflow vulnerability}\label{subsubsec2.1.2}

Identifying test units for stack-based buffer overflow vulnerability is more challenging in comparison with heap-based buffer overflow. This challenge arises since there is no specific instruction in executable codes for defining stack buffers. When a buffer is defined in the high-level source code, no instruction is accordingly added to the corresponding executable code. Therefore, unlike heap buffers, which are explicitly allocated by calling a dynamic memory allocation function, it is not possible to exactly determine the size and location of a stack buffer based on a specific instruction in executable codes.

As a solution, we estimate the location and maximum length of stack buffers according to the calling conventions and the standards of accessing stack variables in executable codes. When a function is called, its arguments, the return address (\texttt{rip}), and the base frame pointer of the caller function (\texttt{rbp}) are respectively pushed into the stack. Then, local variables of the called function are placed into the stack. The local variables of each function are accessed using the base frame pointer of its stack memory, which is stored in a specific register, e.g., \texttt{rbp} in x64 instruction set architecture. In this way, the start address of a stack buffer is calculated as \texttt{rbp-x}, which \texttt{x} helps to estimate the length of that buffer. If some data larger than \texttt{x} bytes is stored in such stack buffer, it overwrites not only other local variables but also the contents of the base frame pointer and probably the return address in the stack. Thus, we locate stack buffers using instructions that access memory buffers with address \texttt{rbp-x}, and estimate their maximum lengths as \texttt{x}.

As an example, Fig.~\ref{fig:fig5} presents the high-level code of a function in C language and its equivalent x64 assembly code. As shown in this figure, the start address of \texttt{buffer} is \texttt{rbp-32}, and thus it is 32 bytes far from the stored base frame pointer in the stack. Therefore, we estimate the maximum length of \texttt{buffer} as 32 bytes.

A drawback of our method in estimating the buffer length is that we can only detect stack overflows that corrupt the \texttt{rbp} value. If stack-based buffer overflow occurs in a way that only local variables within the function are corrupted, but the overflowing data could not be long enough to overwrite the \texttt{rbp} value in the stack, our method is not able to detect the vulnerability. 

Our proposed specification for stack-based buffer overflow vulnerability is presented in Fig.~\ref{fig:fig6}. This specification consists of three events: accessing the \texttt{rbp} register, accessing a stack buffer with address \texttt{rbp-x}, and storing some data in a stack buffer.

\begin{figure}[t]
\vspace{-1.5mm}
\begin{lstlisting}[language=c, xleftmargin=3mm, xrightmargin=1.5mm]
void func()				func: 
{								 push  rbp	
	/*buff definition*/	 mov   rbp, rsp
	char buff[20];			 lea   rax, [rbp-32]
	/*buff usage*/			 mov   DWORD PTR [rax], 7102838
	strcpy(buff, "Val");	 pop   rbp
}								 ret		
\end{lstlisting}
\vspace{3mm}
\caption{An example of a C code and its equivalent assembly code for using stack memory based on x64 architecture\label{fig:fig5}}
\end{figure}

\begin{figure}[t]
    \centering
    \scriptsize
    \fbox{\begin{minipage}{0.46\textwidth}
        \centering
        (\texttt{CONT1},True) $\rhd$ (\{\texttt{CONT2},\texttt{CONT3}\},Rule1) $\rhd$ (\{\texttt{CONT5},\texttt{CONT6}\},Rule2)\\
        \vspace{5pt}
        \hrule
        \begin{enumerate}
            \item \texttt{CONT1 = GET:I64(20)\\
            \hspace{6pt} \textbf{\color{codegreen} Container[SOF].tag = Ist\_WrTmp\\
            \hspace{6pt} Container[SOF].WrTmp.tmp = CONT1\\
            \hspace{6pt} Container[SOF].WrTmp.data->tag = Iex\_Get\\
            \hspace{6pt} Container[SOF].WrTmp.data->Iex.Get.offset = 20\\
            }}
            \item \texttt{CONT4 = Add64(CONT2, CONT3)\\
            \hspace{6pt} \textbf{\color{codegreen} Container[SOF].tag = Ist\_WrTmp\\
            \hspace{6pt} Container[SOF].WrTmp.tmp = CONT4\\
            \hspace{6pt} Container[SOF].WrTmp.data->tag = Iex\_Binop\\
            \hspace{6pt} Container[SOF].WrTmp.data->Iex.Binop.op = Iop\_Add64\\
            \hspace{6pt} Container[SOF].WrTmp.data->Iex.Binop.arg1 = CONT2\\
            \hspace{6pt} Container[SOF].WrTmp.data->Iex.Binop.arg2 = CONT3\\
            }}
            \item \texttt{STle(CONT5) = CONT6\\
            \hspace{6pt} \textbf{\color{codegreen} Container[SOF].tag = Ist\_Store\\
            \hspace{6pt} Container[SOF].Store.addr = CONT5\\
            \hspace{6pt} Container[SOF].Store.data->tag = Iex\_RdTmp\\
            \hspace{6pt} Container[SOF].Store.data->Iex.RdTmp.tmp = CONT6\\
            }}
        \end{enumerate}
        \raggedright Rule1:\\
        \centering
        \texttt{CONT1 $==$ CONT2 $\wedge$ CONT3 $<$ 0}\\
        \raggedright Rule2:\\
        \centering
        \texttt{CONT4 $\leq$ CONT5 $<$ CONT2}
        \texttt{ $\wedge$ }
        \texttt{len(CONT6) $>$ |CONT3|}
        \vspace{1pt}
        \end{minipage}}
    \caption{\centering Specification of stack-based buffer overflow vulnerability in VEX language\label{fig:fig6}}
\end{figure} 

In the first event of the specification, we consider statements in which the program accesses the value of \texttt{rbp} register and stores it in a temporary variable. In this event, the value of \texttt{rbp} register is extracted and stored in a temporary variable using a \texttt{WrTmp} VEX statement. The source and destination of \texttt{WrTmp} are defined by \texttt{WrTmp.data} and \texttt{WrTmp.tmp} parts of this VEX statement, respectively. According to the detailed structure of this statement, \texttt{rbp} register is accessed with a \texttt{WrTmp} statement that its \texttt{WrTmp.data} section is an expression of type \texttt{Get} in which \texttt{Get.offset} is equal to 20. We consider \texttt{WrTmp.tmp} as \texttt{CONT1} to record the temporary variable that stores the value of \texttt{rbp} register.

In the second event, we look for statements that compute the value of \texttt{rbp-x}. This operation is performed in a \texttt{WrTmp} statement in which the \texttt{WrTmp.data} section is a \texttt{Binop} expression of type \texttt{Add64}. The first argument of this operation, which is defined by \texttt{Binop.arg1}, is considered as \texttt{CONT2}, and the second one, defined by \texttt{Binop.arg2}, is considered as \texttt{CONT3}. Rule1 states that the value of \texttt{CONT2} must be equal to the temporary variable in which the content of the \texttt{rbp} register was stored in the previous step (\texttt{CONT1}). Also, \texttt{CONT3} must be a negative number as \texttt{rbp-x} is calculated using the \texttt{Add64} operator. 
The absolute value of the second argument in the \texttt{Binop} expression ($|$\texttt{CONT3}$|$) is assumed as the maximum size of the buffer. The result of this operation is stored in \texttt{WrTmp.tmp} part and is considered as \texttt{CONT4}.

In the third event, we locate \texttt{Store} statements that store some data in a stack buffer. In this statement, \texttt{Store.data} is an expression of type \texttt{RdTmp}, and the temporary variable storing the source data, defined by \texttt{RdTmp.tmp} section, is considered as \texttt{CONT6}. The destination buffer address is also defined by \texttt{Store.addr} part of the \texttt{Store} statement and is considered as \texttt{CONT5}. According to Rule2, stack-based buffer overflow occurs when the destination address in the \texttt{Store} statement (\texttt{CONT5}) is in the range of a stack buffer and the source data (\texttt{CONT6}) is larger than the maximum size of the destination stack buffer.

We use this specification and locate stack buffers throughout the program's executable code and consider the functions in which a store operation is performed on a stack buffer as test units. It is worth mentioning that we use the same method as in the previous section to trace the stack buffers in nested function calls.

\subsubsection{Test unit of double-free vulnerability}\label{subsubsec2.1.3}

Our proposed specification for double-free vulnerability is presented in Fig.~\ref{fig:fig7}. This vulnerability arises in a scenario with an allocation of a heap buffer and two times freeing of that buffer. 

\begin{figure}[t]
    \centering
    \scriptsize
    \fbox{\begin{minipage}{0.46\textwidth}
        \centering
        (\{\texttt{CONT1},\texttt{CONT2}\},True) $\rhd$ (\texttt{CONT3},Rule1) $\rhd$ (\texttt{CONT4},Rule2)\\
        \vspace{5pt}
        \hrule
        \begin{enumerate}
            \item \texttt{CONT1 = malloc(CONT2)}
            \item \texttt{free(CONT3)}
            \item \texttt{free(CONT4)}
        \end{enumerate}
        \raggedright Rule1:\\
        \centering
        \texttt{CONT1 $\leq$ CONT3 $<$ CONT1 $+$ CONT2}\\

        \raggedright Rule2:\\
        \centering
        \texttt{CONT1 $\leq$ CONT4 $<$ CONT1 $+$ CONT2}\\
        \vspace{1pt}
        \end{minipage}}
    \caption{Specification of double-free vulnerability in VEX\label{fig:fig7}}
\end{figure}

\begin{figure}[t]
%\vspace{-1.5mm}
\begin{lstlisting}[language=c, xrightmargin=1.5mm]
char * child1(char * source)
{
	source = (char *)malloc(10*sizeof(char));
	return source;
}
void child2(char * source, int n)
{
	char * src_copy;
	src_copy = (char *)malloc(n*sizeof(char));
	strncpy(src_copy, source, n);
	free(source);
	/* some lines of code */
	free(src_copy);
}
void parent()
{
	int n = 10; char * source;
	source = child1(source);
	child2(source, n);
	/* FLAW: freeing memory twice */
	free(source);
}
int main(void)
{
	parent(); return 0;
}
\end{lstlisting}
\vspace{3mm}
\caption{Example of a C code with double-free vulnerability\label{fig:fig8}}
\end{figure}

For the first event, we locate all \texttt{malloc} function calls to determine the addresses of pointers that point to the allocated buffers and consider them as \texttt{CONT1}. For the second and third events, we find all \texttt{free} function calls in the program's executable code, and extract their input argument demonstrating the pointer address of the released buffer as \texttt{CONT3} and \texttt{CONT4}. According to Rule1 and Rule2, double-free vulnerability occurs when the memory address of a heap buffer is freed by calling the \texttt{free} function twice.

The test unit for double-free vulnerability is identified as a function that consists of the three events related to this specification, i.e., buffer allocation, buffer deletion, and buffer re-deletion. 
As an example, in the sample C code in Fig.~\ref{fig:fig8}, the memory allocation operation is performed in function \texttt{child1}, then the allocated heap buffer is deleted in function \texttt{child2}, and again deleted in function \texttt{parent}. In this example, \texttt{parent} is the function that contains all three events of the vulnerability specification, which happen sequentially. Hence, \texttt{parent} is considered a test unit for double-free vulnerability in this code.

\subsubsection{Test unit of use-after-free vulnerability}\label{subsubsec2.1.4}

Our proposed specification for use-after-free vulnerability is presented in Fig.~\ref{fig:fig9}. This vulnerability appears in a scenario with three events: allocating a heap buffer, deleting the same buffer, and then using that buffer in a read operation with a \texttt{WrTmp} statement or in a store operation with a \texttt{Store} statement. The \texttt{Not\_Imp} value in this specification is assigned to the parts of the statements, which are not important.

%More specifically, use-after-free occurs when the freed memory is not reallocated before getting used. This is mentioned in the third part of the specification, which means there should be no pair (\texttt{CONT4},Rule2) in the event sequence of this vulnerability.

\begin{figure}[t]
    \centering
    \scriptsize
    \fbox{\begin{minipage}{0.46\textwidth}
        \centering
        (\{\texttt{CONT1},\texttt{CONT2}\},True) $\rhd$ (\texttt{CONT3},Rule1) $\rhd$ ((\texttt{CONT4},Rule2) $|$ (\texttt{CONT5},Rule3))\\
        \vspace{5pt}
        \hrule
        \begin{enumerate}
            \item \texttt{CONT1 = malloc(CONT2)}
            \item \texttt{free(CONT3)}
            \item \texttt{Not\_Imp = LDle:I64(CONT4)\\
            \hspace{6pt} \textbf{\color{codegreen} Container[UAF].tag = Ist\_WrTmp\\
            \hspace{6pt} Container[UAF].WrTmp.tmp = Not\_Imp\\
            \hspace{6pt} Container[UAF].WrTmp.data->tag = Iex\_Load\\
            \hspace{6pt} Container[UAF].WrTmp.data->Iex.Load.addr = CONT4\\
            }}
            \item \texttt{STle(CONT5) = Not\_Imp\\
            \hspace{6pt} \textbf{\color{codegreen} Cont[UAF].tag = Ist\_Store\\
            \hspace{6pt} Container[UAF].Store.addr = CONT5\\
            \hspace{6pt} Container[UAF].Store.data->tag = Iex\_RdTmp\\
            \hspace{6pt} Container[UAF].Store.data->Iex.RdTmp.tmp = Not\_Imp\\
            }}
        \end{enumerate}
        \raggedright Rule1:\\
        \centering
        \texttt{CONT1 $\leq$ CONT3 $<$ CONT1 $+$ CONT2}\\
        \raggedright Rule2:\\
        \centering
        \texttt{CONT1 $\leq$ CONT4 $<$ CONT1 $+$ CONT2}\\
        \raggedright Rule3:\\
        \centering
        \texttt{CONT1 $\leq$ CONT5 $<$ CONT1 $+$ CONT2}\\
        \vspace{1pt}
        \end{minipage}}
    \caption{Specification of use-after-free vulnerability in VEX\label{fig:fig9}}
   
\end{figure}

According to this specification, we consider functions that contain \texttt{malloc}, \texttt{free}, and load or store operations on the same heap buffer as a test unit.

\subsection{Processing the test units}\label{subsec2.2}

In this phase, we perform symbolic execution for the extracted test units. For each test unit, a constraint tree is generated that represents basic blocks of the program as its nodes. We annotate each node with some metadata that indicates the system state at that point of the program and contains the path constraints from the beginning of the test unit to the given node, the node constraints, and vulnerability constraints. In the following, we use $Term(n)$, $Const(n)$, and $VulConst(n)$ to represent the node constraints, the path constraints from the beginning of the test unit to the given node $n$, and the vulnerability constraints of node $n$, respectively.

\begin{figure*}[t]
%\vspace{-2em}
\begin{subfigure}{0.46\textwidth}
\centering
\includegraphics[width=0.8\textwidth]{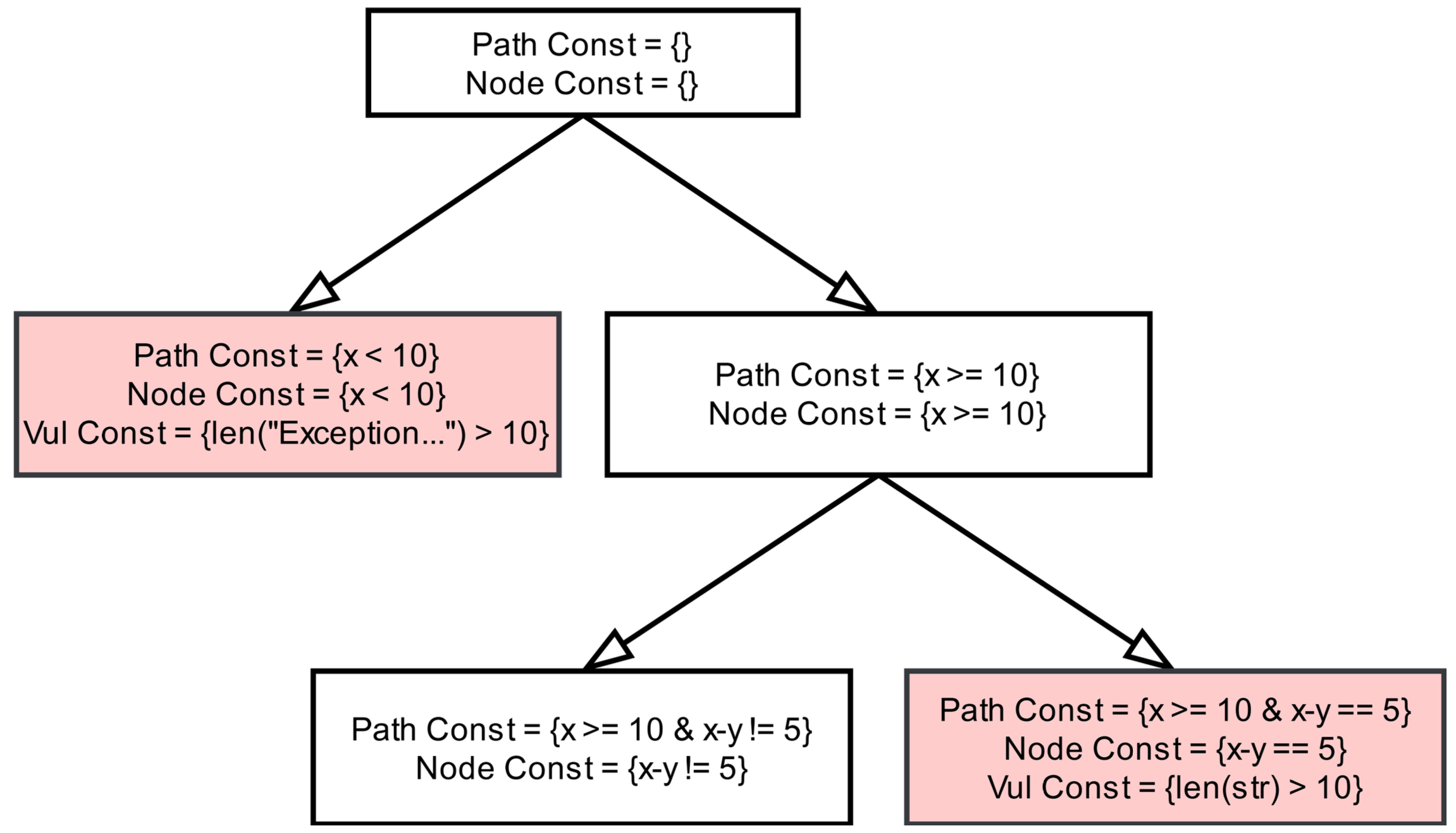}
\caption{Extracted unit tree\label{fig:first}}
\end{subfigure}
\hfill
\begin{subfigure}{0.475\textwidth}
%\vspace{1.5cm}
\begin{lstlisting}[language=c, xrightmargin=1mm]
void testUnit(int x, int y, char * str) 
{
	char * msg = (char *)malloc(10);
	if(x >= 10) {
		if (x-y == 5) {
			strcpy(msg, str);
		}
	}
	else {
		strcpy(msg, "Exception...");
	}
	printf("%s", msg);
}
\end{lstlisting}
\hfill
\vspace{0.5em}
\caption{Source code of a test unit}
\label{fig:second}
\end{subfigure}
\hfill
\caption{An example of a test unit and its corresponding constraint tree\label{fig:fig10}}
\end{figure*}

The vulnerability constraints are calculated according to the vulnerability specifications in the previous section. To make this step clear, an example test unit and its corresponding constraint tree are presented in figure Fig.~\ref{fig:fig10}. In this figure, the two red nodes represent basic blocks with probably vulnerable statements because they copy some data into the \texttt{msg} buffer. Thus, according to size of the \texttt{msg} buffer, a vulnerability constraint is calculated in addition to the path constraints for these nodes. 

\subsection{Learning and simulation process}\label{subsec2.3}

After extracting the constraint tree, the program execution is simulated, and its behavior is learned in the third phase of our solution. Details of the operations in this phase are presented in Algorithm~\ref{alg:alg1}. This algorithm is a revised version of the \textit{Cover} algorithm presented in \cite{refbib19}, and our modifications are shown in blue.

In this algorithm, first in lines 1 and 2, we perform $n$-factor Monte Carlo simulation on the program by picking values over a $d$-dimensional space for $d$ input arguments. We generate possible combinations of input vectors and execute the program by giving them as system arguments. For each system input vector, we monitor the program execution and record the corresponding unit input vector to annotate nodes of the unit tree with pairs of the system and correlated unit input vectors $(V_k, v_k)$ that reach these nodes. 

Next, in lines 3 and 4, we explore the constraint tree and analyze only nodes located in a possibly vulnerable path. A vulnerable path is defined as a path from the root to a leaf in the constraint tree, which contains some nodes with probably vulnerable statements, e.g., a store operation to a stack buffer or a \texttt{free} function call. In contrast to the algorithm in \cite{refbib19}, which processes all nodes of the constraint tree at this step, we restrict our analysis to fewer nodes according to the specification of the target vulnerability class to improve the efficiency of our method. 

In lines 5 to 9, we check if these nodes and their siblings have been executed during the simulation. If yes, we use \textit{TAR3} treatment learning algorithm \cite{refbib32} to estimate the range of system inputs that could explore the desired node in the unit. Otherwise, in lines 10 to 12, for each uncovered node whose path constraints are satisfiable, a function named \textit{ComputeMap} is called that estimates the correlation between system and unit input data as a function $f_n$. The algorithm of this function is presented in Algorithm~\ref{alg:alg2}.

\subsection{Test data generation}\label{subsec2.4}

\begin{algorithm}[t]
    \footnotesize
    \caption{Cover$(S, U,\color{blue} T \color{black})$\label{alg:alg1}}
    \begin{algorithmic}[1]
    \Require {System $S$ with inputs $I$ with $d=|I|$, unit $U$ with inputs $i$ and \color{blue} constraint tree $T$ obtained from applying symbolic execution to the unit $U$}
    \color{black}
    \State {Perform $n$-factor combinatorial MC simulation over space $R^d$}
    \State {$(V,v) \gets \{(a,b) |$ $a$ is a system level vector and $b$ is the corresponding monitored unit level vector$\}$}
    \For{node $n$ in $T$ using BFS}
        \color{blue}
        \If{$n$ is in a possibly vulnerable path}
        \color{black}
            \If{$n$ and $n$’siblings are covered} 
                \State {$V' \gets \{a \in V |$ $a$ cover $ n\}$} 
                \State {$V'' \gets V \backslash V'$}
                \State {$(I_n,R_n,\_) \gets $ RunTar3$(I,V,V',V'')$}
                \State {$\forall j \in I_n$ store the range $r_j \in R_n$ for $j$}
            \ElsIf{$n$ is satisfiable but not covered}
                \State {$C \gets Term(n)$}
                \State{$(I_n,i_n,f_n) \gets $ ComputeMap$(C, I, V, v, n, Parent(n))$}
            \EndIf
        \color{blue}
        \EndIf
        \color{black}
    \EndFor
    \color{blue}
    \For{node $n$ in $T$ that $n$ is possibly vulnerable and satisfiable}
       %\If{$n$ is vulnerable to stack overflow or heap overflow} 
       		\State {$C_n \gets$ $Const(n) \wedge VulConst(n)$}
       %\ElsIf{$n$ is vulnerable to double-free or use-after-free}
       %		\State {$C_n \gets$ $Const(n)$}
       %\EndIf
       \If{$n$ is covered}  
            \State {Generate input using $C_n$ and $\forall j \in I_n$ use $r_j$ from line 9}
        \Else
            \State {Generate input using $C_n$ and $f_n$ from line 12}
        \EndIf
    \EndFor
    \end{algorithmic}
\end{algorithm}

Until now, we have only considered path constraints in generating system input data. In lines 16 to 23 of Algorithm ~\ref{alg:alg1}, for each node containing a possibly vulnerable statement whose path constraints are satisfiable, we attempt to generate system input data consistent with both path and its calculated vulnerability constraints. To do so, in lines 18 and 19, for each node that has been covered in the simulation step, we solve the path and vulnerability constraints of the node and generate appropriate unit input data to reveal vulnerability in that node. In the last step, using the range of system inputs calculated by \textit{TAR3} algorithm, we find relevant system input data for the intended unit input data.

Next, in lines 20 and 21, for each node that has not been covered in the simulation step, we use the fitted function $f_n$ to find relevant system input data for the unit input data consistent with calculated path and vulnerability constraints.

To summarize the difference between our \textit{Cover} algorithm and the one presented in \cite{refbib19}, first in line 4, we improve the performance of our analysis by only considering nodes in potentially vulnerable paths, while in \cite{refbib19}, all the nodes are analyzed in this step even though they might not contain any vulnerability. Next, we consider both path and vulnerability constraints, and this is statically performed using symbolic execution. However, the algorithm in \cite{refbib19} only considers the path constraints calculated gradually using dynamic symbolic execution by generating new input data that explore uncovered paths in the unit. Thus, we calculate the constraints more quickly. Since our symbolic analysis is restricted to a single function, dynamic symbolic execution accuracy and coverage advantages over symbolic execution are not significant here. Finally, we consider the path and vulnerability constraints, in line 17, to generate appropriate system input values that reach the intended nodes in the test unit and activate their vulnerabilities. In contrast, the algorithm in \cite{refbib19} only considers the path constraints for generating system inputs that cover the nodes of the unit.

\subsubsection{ComputeMap algorithm}\label{subsubsec2.4.1}

We have used the same algorithm as introduced in \cite{refbib19} for the \textit{ComputeMap} function, which is presented in Algorithm~\ref{alg:alg2}. Here, we describe this algorithm to clarify the whole process for the reader. In summary, this algorithm attempts to define the correlation of system and unit input parameters as a function. Due to the complexity of applying curve fitting to a large set of data and the presence of a large number of parameters, the algorithm initially considers the constraints of each node individually. More precisely, instead of considering all parameters in $Vars(Const(n))$ as $i_n$, the algorithm first attempts to reduce the unit and system input parameters by selecting a subset of unit input parameters $i_n$ appearing in $Vars(Term(n))$ and a subset of system input parameters $I_n$ that are most effective on the values of parameters in $i_n$.

Thus, the unit input parameters related to the node constraints are extracted, and the first 20\% of system vectors that are more compatible with this constraints subset are selected as a set $V'$. Afterwards, \textit{TAR3} algorithm is applied to the sets $V' \subset V$ and $V'' = V \textbackslash V'$, and if a smooth relationship is established therebetween, the function $f_n$ is built using curve fitting. Otherwise, the process is recursively repeated by adding parent node constraints to the given node in order to establish a smooth relationship. A smooth function is a function that has continuous derivatives up to a specific order.

If a smooth relationship is not found by including all terms in $Const(n)$, in lines 13 to 20, we walk up through the unit tree to find a parent node with enough system input vectors in its annotation. Such a node is covered in the simulation step with an appropriate number of input vectors $(V_k, v_k)$ that helps to better estimate the function $f_n$ using the curve fitting algorithm.

\begin{algorithm}[t]
    \footnotesize
    \caption{ComputeMap$(C,I,V,v,n,n')$\label{alg:alg2}}
    \begin{algorithmic}[1]
    \Require {Constraints set $C$, System Inputs $I$, System vectors $V$, Unit vectors $v$, a node $n$ that we want to cover, a node $n'$ that is in the parent hierarchy of $n$ }
    \Ensure {$(I_n,i_n,f_n)$ where $i_n = Vars(C)$ and $I_n = f(i_n)$}
    \State {$i_n = Vars(C)$}
    \State {$V' \gets \{a \in V |$ $a$ is in 20\% of points closet to $C \}$}
    \State {$V'' \gets V \backslash V'$}
    \State {$(I_n,R,Smooth) \gets$ RunTar3$(I,V,V',V'')$}
    \If{$Smooth$}
        \State {Build map $I_n = f(i_n)$ \Comment{curve fitting step}}
    \Else
        \If{$n'$ exists} 
            \State {$C \gets C \wedge Term(n')$}
            \State {$(I_n,i_n,f_n) \gets$ ComputeMap$(C,I,V,v,n,Parent(n'))$}
        \Else 
            \State {$n'' \gets n$}
            \While{$Parent(n'')$ exists}
                \State {$C \gets C \wedge Term(Parent(n''))$}
                \State {$n'' \gets Parent(n'')$}
                \State {$V' \gets \{a \in V |$ $a$ cover $n''\}$}
                \If{$|V'| \geq Threshold$ }
                    \State {break}
                \EndIf
            \EndWhile
            \State {$V'' \gets V \backslash V'$}
            \State {$(I_n,R_n,\_) \gets$ RunTar3$(I,V,V',V'')$}
            \State {$i_n = Vars(C)$}
            \State {Build map $I_n = f(i_n)$ \Comment{curve fitting step}}
        \EndIf
    \EndIf
    \end{algorithmic}
\end{algorithm}
%\vspace{-1em}

\section{Evaluation}\label{sec3}
\textit{UbSym} is implemented as a plugin for \textit{angr} framework. In our implementation, \texttt{string} data type is also supported for detecting vulnerabilities in string manipulation functions in addition to \texttt{int}, \texttt{short}, \texttt{unsigned int}, \texttt{char}, \texttt{float}, \texttt{double}, and \texttt{enum} data types supported in the proposed approach in \cite{refbib19}.

We have designed two experiments to evaluate our solution. In the first experiment, a set of 225 benchmark programs provided by the National Institute of Standards and Technology in Software Assurance Reference Dataset (SARD) project \cite{refbib30} is selected. This set is divided into four groups of vulnerable programs for evaluating the detection performance of four vulnerability classes. The test programs in this set contain a wide range of vulnerable instructions, which helps us to examine our method in various scenarios.

The test programs in NIST SARD are classified based on the classification of vulnerabilities in the Common Weakness Enumeration (CWE) database. We have tested our tool on the programs of classes \seqsplit{CWE122\_Heap\_Based\_Buffer\_Overflow} and \seqsplit{CWE121\_Stack\_Based\_Buffer\_Overflow} for heap-based and stack-based buffer overflow vulnerabilities. The vulnerability occurs in these programs when some constant data is copied into a heap or stack buffer using \texttt{strcpy}, \texttt{strcat}, \texttt{memcpy}, and \texttt{memmove} functions. To better evaluate our proposed method, we have made the path constraints in the test programs more complicated by adding an additional \texttt{if} statement to the vulnerable paths. In addition, instead of copying constant data into a heap or stack buffer, we have copied an input variable, entered by the user as a command-line argument, into that buffer to create a vulnerability constraint in the test unit. A vulnerable function in one of these benchmark programs is presented in Fig.~\ref{fig:fig11} as an example and our added \texttt{if} statement is underlined in line 15. The same \texttt{if} statement is similarly added to all benchmark programs.

\begin{figure*}[t]
\begin{lstlisting}[language=c, xrightmargin=1mm]
void CWE121_Stack_Based_Buffer_Overflow__CWE805_char_declare_memcpy_34_bad(char * source)
{
    char * data;
    CWE121_Stack_Based_Buffer_Overflow__CWE805_char_declare_memcpy_34_unionType myUnion;
    char dataBadBuffer[50];
    char dataGoodBuffer[100];
    /* FLAW: Set a pointer to a "small" buffer. This buffer will be used in the sinks as a destination * buffer in various memory copying functions using a "large" source buffer. */
    data = dataBadBuffer;
    data[0] = '\0'; /* null terminate */
    myUnion.unionFirst = data;
    {
        char * data = myUnion.unionSecond;
        {
            /* POTENTIAL FLAW: Possible buffer overflow if the size of data is less than the length of source */
            ifä\underline{(source[0] == '7' \&\& source[1] == '/' \&\& source[2] == '4' \&\& source[3] == '2' \&\& source[4] == 'a' \&\& source[5] == '8' \&\& source[75] == 'a')}ä {
                memcpy(data, source, strlen(source)*sizeof(char));
            }
            data[100-1] = '\0'; /* Ensure the destination buffer is null terminated */
            printLine(data);
        }
    }
}
\end{lstlisting}
\vspace{3mm}
\caption{A sample vulnerable unit in benchmark programs\label{fig:fig11}}
\end{figure*}

We have also tested our tool on CWE416\_Use\_After\_Free and CWE415\_Double\_Free benchmark programs to evaluate the performance of our method in detecting use-after-free and double-free vulnerabilities, respectively. The \texttt{if} statement mentioned above is also added to the vulnerable path in these programs. In \seqsplit{CWE415\_Double\_Free} programs, the \texttt{if} statement is placed on top of the second \texttt{free} function call and in \seqsplit{CWE416\_Use\_After\_Free} programs, it is placed on top of the memory usage instruction.

In the second experiment, we have created a test program for each vulnerability class with several functions and more complicated path and vulnerability constraints to better evaluate the efficiency of our method. The source code of these programs, along with their details, are presented in the Appendix.

In both experiments, we have compared \textit{UbSym} with two other tools, MACKE \cite{refbib24} and Driller \cite{refbib25}, which use similar methods for detecting various vulnerability classes in C programs.

\begin{table*}[!t]
\caption{Results of evaluating the approaches on the selected group of test programs in NIST benchmark programs\label{tab1}}
{\begin{tabular*}{\textwidth}{@{\extracolsep{\fill}}lcccccccccccc@{}}\toprule

\multicolumn{1}{c}{} & \multicolumn{3}{c}{Heap\_Based\_Buffer\_Overflow} & \multicolumn{3}{c}{Stack\_Based\_Buffer\_Overflow} & \multicolumn{3}{c}{
Use\_After\_Free} & \multicolumn{3}{c}{Double\_Free} \\
\cmidrule{2-4} \cmidrule{5-7} \cmidrule{8-10} \cmidrule{11-13}
Method & Acc. & Prec. & Rec. & Acc. & Prec. & Rec. & Acc. & Prec. & Rec. & Acc. & Prec. & Rec. \\
\midrule
MACKE & 0.78 & 0.87 & 0.60 & 0.81 & 0.75 & 0.84 & 0.21 & 1.00 & 0.21 & 0.82 & 1.00 & 0.82\\
Driller & 1.00 & 1.00 & 1.00 & 0.90 & 1.00 & 0.77 & - & - & - & 0.85 & 1.00 & 0.85\\
\textit{UbSym} & 1.00 & 1.00 & 1.00 & 1.00 & 1.00 & 1.00 & 1.00 & 1.00 & 1.00 & 1.00 & 1.00 & 1.00\\
\bottomrule
\end{tabular*}}{}
\end{table*}

\subsection{Experiment 1}\label{subsec3.1}

\begin{figure}[t]
\begin{equation}\label{eqn2}
Precision = \frac{TP}{TP + FP}
\end{equation}
\begin{equation}\label{eqn3}
Recall = \frac{TP}{TP + FN}
\end{equation}
\begin{equation}\label{eqn4}
Accuracy = \frac{TP + TN}{TP + TN + FP + FN}
\vspace{0.25cm}
\end{equation}
    \centering
    \begin{tikzpicture}
    \begin{axis}
    [  
        ybar,  
        symbolic x coords={CWE122,CWE121,CWE416,CWE415},
        nodes near coords,  
        nodes near coords align={vertical},  
        every node near coord/.append style={font=\tiny, rotate=90, xshift=3mm, yshift=-2mm},
        xtick=data,  
        xticklabel style={
        font=\footnotesize,
        align=right,
        rotate=0,
        },
        yticklabel style={
        font=\footnotesize,
        },
        ymin=0, ymax=800,
        ylabel={\footnotesize Execution Time (s)},  
        width=0.45\textwidth,
        height=6cm,
        bar width=8pt,
        legend style={at={(1,1)},anchor=north east}
    ]  
    \addplot[fill=lightgray] coordinates {
    (CWE122,46) 
    (CWE121,77) 
    (CWE416,73) 
    (CWE415,16) 
    };  

    \addplot [fill=black] coordinates {
    (CWE122,433) 
    (CWE121,694) 
    (CWE416,0) 
    (CWE415,297)  
    }; 
    
    \addplot [fill=gray] coordinates { 
    (CWE122,18) 
    (CWE121,37) 
    (CWE416,14) 
    (CWE415,10)
    };   
    \legend{\footnotesize{UbSym}, \footnotesize{Driller}, \footnotesize{MACKE}}  
    \end{axis}  
    \end{tikzpicture}
    \caption{The average analysis time of the tools for a group of test programs in NIST benchmark programs\label{fig:fig12}}
    %\vspace{-1em}
\end{figure}

Table~\ref{tab1} represents the results of our first experiment in testing NIST SARD vulnerable programs. Each test program in this set has a function whose name includes the word \texttt{bad}, which contains a vulnerable statement, and one or multiple functions whose names include the word \texttt{good}, containing similar statements without vulnerability. Thus, we expect to achieve exactly one true positive alarm for each test program using a precise vulnerability detection mechanism. We have used three evaluation metrics in this experiment, i.e., Precision, Recall, and Accuracy, as shown in (\ref{eqn2}), (\ref{eqn3}), and (\ref{eqn4}), respectively. In these equations, TP is the number of true positives, TN is the number of true negatives, FP is the number of false positives, and FN is the number of false negatives. Since \textit{UbSym} detects vulnerabilities in the test programs of this experiment in less than 120 seconds, we have set the timeout value as 900 seconds. Thus, if a tool does not detect a vulnerability in a test program in less than 900 seconds, we consider it a false negative.

\begin{table*}[t]
\caption{Results of evaluating the approaches on the complicated test programs\label{tab2}}
{\begin{tabular*}{\textwidth}{@{\extracolsep{\fill}}lcccccccccccccccccccc@{}}\toprule

\multicolumn{1}{c}{} & \multicolumn{5}{c}{Heap\_Based\_Buffer\_Overflow} & \multicolumn{5}{c}{Stack\_Based\_Buffer\_Overflow} & \multicolumn{5}{c}{
Use\_After\_Free} & \multicolumn{5}{c}{Double\_Free} \\
\cmidrule{2-6} \cmidrule{7-11} \cmidrule{12-16} \cmidrule{17-21}
Method & TP & FP & TN & FN & Time,s & TP & FP & TN & FN & Time,s & TP & FP & TN & FN & Time,s & TP & FP & TN & FN & Time,s \\
\midrule
MACKE & 3 & 0 & 0 & 3 & 290 & 3 & 0 & 0 & 3 & 193 & 1 & 0 & 0 & 3 & 141 & 0 & 0 & 0 & 4 & 140\\
Driller & 4 & 0 & 0 & 2 & 3325 & 4 & 0 & 0 & 2 & 3433 & - & - & - & - & - & 3 & 0 & 0 & 1 & 3425\\
\textit{UbSym} & 6 & 0 & 0 & 0 & 517 & 6 & 0 & 0 & 0 & 1340 & 4 & 0 & 0 & 0 & 110 & 4 & 0 & 0 & 0 & 88\\
\bottomrule
\end{tabular*}}{}
\end{table*}

Given the experimental results in Table~\ref{tab1}, \textit{UbSym} has detected all vulnerabilities of NIST SARD test programs with no false alarms. Moreover, it has discovered
523 test units in the first step, when applying static analysis on the benchmark programs, among which only 225 test units are actually vulnerable. This way, \textit{UbSym} has omitted 298 units after applying symbolic execution as it could not find any vulnerable node, in the corresponding unit trees, whose path and vulnerability constraints are satisfiable.

Considering the average execution time of analyzing each group of test programs, as shown in Fig.~\ref{fig:fig12}, \textit{UbSym} was considerably faster than Driller. Although the analysis time of MACKE in this experiment has been less than that of our tool, it has generated more false alarms and less accurate results. Additionally, MACKE only generates local input data for executing a single unit and does not consider the path constraints out of the unit. On the contrary, our proposed method generates accurate test data values for running the whole program from the beginning and reaching the vulnerable statement in the test unit. This is a reason that causes our method to take more time to test and analyze a program.

It is worth mentioning that Driller only detects vulnerabilities making the program crash. Therefore, it has not been tested on \seqsplit{CWE416\_Use\_After\_Free} programs in our experiments as it could not detect vulnerability in these programs. To be more precise, in
Fig.~\ref{fig:fig12}, about 28\% of the total analysis time in analyzing CWE122-Heap-Based-Buffer-Overflow programs refers to the execution of the Cover algorithm and curve fitting process. Also, for analyzing CWE121-Stack-Based-Buffer-Overflow programs, 33\% of the analysis time and for analyzing CWE415-Double-Free and CWE416-Use-After-Free programs, 56\% of the total time refers to the learning process, and the rest of
it is spent to apply symbolic execution. 

\subsection{Experiment 2}\label{subsec3.2}

In the second experiment, we have designed four test programs that contain more vulnerable instructions and more complicated constraints in comparison with the test programs in the first experiment. The structures of these programs are similar, but they have different vulnerable statements. There are six vulnerability occurrences through three functions in programs that contain stack-based or heap-based buffer overflow and four vulnerability occurrences through two functions in programs that contain use-after-free or double-free vulnerability. Also, there are various path constraints inside and outside of each test unit. Thus, \textit{UbSym} is supposed to identify test units and analyze them to generate twenty true positive alarms in the test programs altogether. The source code of the test programs in this experiment, along with its details, is presented in the Appendix.

The results of this experiment are demonstrated in Table~\ref{tab2}. In this experiment, we have set the timeout value for analyzing each program as 1 hour. As shown in Table~\ref{tab2}, MACKE could detect only seven vulnerability occurrences in the given programs as it could not analyze complicated path conditions. As well, Driller could only detect eleven vulnerability occurrences since it takes much time to analyze all execution paths. On the contrary, \textit{UbSym} has been able to precisely detect all vulnerabilities and it has generated appropriate test data for the whole program, which enter the programs from the beginning and reveal the vulnerability in the test units. Besides, the testing time for \textit{UbSym} has been much less than that of Driller.

The results of these two experiments clearly demonstrate the advantage of restricting symbolic execution scope and applying machine learning techniques in estimating the program's behavior for detecting various vulnerability classes. This way, we leverage symbolic execution and machine learning techniques in a complementary manner to elevate the efficiency of vulnerability detection.  

\section{Discussion}
\label{diss}
In this section, we discuss the limitations of \textit{UbSym} and some future directions to further improve automated vulnerability detection. 

In the current version of \textit{UbSym}, we restrict the scope of symbolic execution and decrease the number of symbolic variables by only considering function (test unit) parameters. This way, we tackle the path explosion problem in a large group of programs. Though by limiting symbolic execution to test units we significantly lower the chance of path explosion, in extremely large units \textit{UbSym} might fail to generate the unit tree for an entire test unit. Thus, we intend to use some pruning techniques to elevate the scalability of our method in the future. 

Another limitation of \textit{UbSym} is the case when stack-based buffer overflow occurs in a way that only local variables inside the test unit are corrupted. Since we approximate the maximum size of stack buffers, \textit{UbSym} could only detect vulnerabilities in cases where the content of the base frame pointer could be overwritten by a sufficiently long data. 

Another interesting future extension of our method might be to enhance the accuracy and time of the learning process by employing new learning models, which in turn improves the overall process of system input generation from unit inputs.

\section{Conclusion}\label{sec4}

While symbolic execution is sound and complete in theory, this method faces challenges in testing real-world programs, such as path explosion. The number of symbolic states grows exponentially, and thorough analysis of real-world programs becomes infeasible.

We proposed a method for applying symbolic execution to detecting four classes of memory corruption vulnerabilities in executable codes, i.e., heap-based buffer overflow, stack-based buffer overflow, use-after-free, and double-free. We limit the scope of symbolic execution to the test units to avoid path explosion. We specified intended memory corruption vulnerability classes in executable codes and presented a method for automatically determining the test units in arbitrary programs accordingly. Using symbolic execution, we generate appropriate unit input data for detecting memory corruption vulnerabilities according to the path and vulnerability constraints in vulnerable statements of the test unit. Then, we use machine learning techniques to estimate the relation between system and unit input data as a function and find consistent system input data that enter into the program from the beginning, cause the execution of vulnerable statements in the test unit, and reveal their vulnerabilities. The experiments showed that this method achieves more efficient and accurate results in detecting vulnerabilities in complex programs compared to similar tools.

%Nothing should interrupt the references.

\balance 
\section*{Appendix}\label{sec5}

The base structure of the designed complex programs is a simple authentication code by which users carry out sign-up and sign-in operations. The source code of these programs is presented in Fig.~\ref{fig:appendix}. To generate the test program of each vulnerability class, the commented lines for each specific vulnerability should be uncommented. In the following, the structure of the test program that contains heap-based buffer overflow vulnerability is explained as an instance.

This program begins by receiving a username and password in the console to sign-up a user. If the condition in line 99 is satisfied, the vulnerable function \texttt{signup} would be called. In this function, two heap buffers are allocated in lines 6 and 7. As there are two copy operations with \texttt{memcpy} function calls in lines 14 and 17, our solution identifies this function as a test unit. There is a path constraint in this function in line 10; therefore, if the input strings for username and password satisfy the path constraints in lines 99 and 10, and their lengths are more than the lengths of the destination heap buffers in the copy operations, they would cause heap-based buffer overflow. Note that the path constraint in line 99 is out of the test unit and should be determined through machine learning. \textit{UbSym} calculates the path constraints in line 10 using symbolic execution. It generates appropriate input data for the \texttt{scanf} operations in line 95, which are consistent with both path constraints inside and outside the unit.

There are two other test units in this program, \texttt{authentication} and \texttt{check} functions, which cause heap-based buffer overflow by calling \texttt{memcpy} and \texttt{strcpy} functions, respectively. The same challenge exists in these functions for our solution to calculate the path constraints inside the test unit and estimate the ones outside it. \textit{UbSym} could successfully identify these units and generate appropriate test data for the whole program, which explore vulnerable instructions in the unit and cause heap-based buffer overflow.% in them.
% *************************************************
\vspace{1em}
\begin{figure*}[!h]
\vspace{-1em}
\begin{lstlisting}[language=c, xrightmargin=1mm]
#define def_user "u-admin-"
#define def_pass "password"
void signup(char *username, char *password)
{
	/* Uncomment the following two lines for heap-based buffer overflow & use-after-free & double-free */
	// char* tmp_user = (char *)(malloc(50*sizeof(char)));
	// char* tmp_pass = (char *)(malloc(50*sizeof(char)));
	/* Uncomment the following line for stack-based buffer overflow */
	// char tmp_user[16], tmp_pass[16];
	if((username[1] >= '0' && username[1] <= '9') && !strncmp(password, "passW0rd", 8))
	{ 
		/* POTENTIAL FLAW: data may not have enough space to hold source */
		/* Uncomment the following line for heap-based & stack-based buffer overflow */
		// memcpy(tmp_user, username, strlen(username));
		/* POTENTIAL FLAW: data may not have enough space to hold source */
		/* Uncomment the following line for heap-based & stack-based buffer overflow */
		// memcpy(tmp_pass, password, strlen(password));
		if(strlen(tmp_pass) < 12) { printf("The selected password is too weak\n"); return; }
		int fd = open(tmp_user, O_WRONLY|O_CREAT, 0777);
		if(fd < 0) { printf("An unexpected problem occurred!\n"); return; }
		write(fd,tmp_pass, sizeof(tmp_pass));
		printf("%s your registration was successful\n", tmp_user);
		/* POTENTIAL FLAW: Free data here - line 30 frees data as well */
 		/* Uncomment the following line for double-free */
		// free(tmp_user); free(tmp_pass);
	}
	else if(!(username[1] >= '0' && username[1] <= '9')) {printf("The second letter of username must be a number\n");}
	else { printf("The password must start with the word <passW0rd>\n"); }
	/* Uncomment the following line for double-free & use-after-free & heap-based buffer overflow */
	// free(tmp_user);	free(tmp_pass);
	/* POTENTIAL FLAW: Use of data that may have been freed in line 30 */ 
	/* Uncomment the following line for use-after-free */
	// tmp_user[50-1] = '\0'; tmp_pass[50-1] = '\0';
}
bool check(char *username, char *password)
{
	/* Uncomment the following two lines for heap-based buffer overflow & use-after-free & double-free */
	// char* tmp_user = (char *)(malloc(50*sizeof(char)));
	// char* tmp_pass = (char *)(malloc(50*sizeof(char)));
	/* Uncomment the following line for stack-based buffer overflow */
	// char tmp_user[16], tmp_pass[16];
	if((username[0] >= 'A' && username[0] <= 'Z') && (username[1] >= '0' && username[1] <= '9'))
	{
		/* POTENTIAL FLAW: data may not have enough space to hold source */
		/* Uncomment the following line for heap-based & stack-based buffer overflow */	    
		// strcpy(tmp_user, username);
		/* POTENTIAL FLAW: data may not have enough space to hold source */
		/* Uncomment the following line for heap-based & stack-based buffer overflow */	    
		// strcpy(tmp_pass, password);
		if(!strcmp(tmp_user, def_user) && !strcmp(tmp_pass, def_pass)) {return true;}
		char passwd[50]; int fd = open(tmp_user, O_RDONLY);
		if(fd < 0) { printf("An unexpected problem occurred!\n"); return false; }
		read(fd, passwd, sizeof(passwd));
		if(!strcmp(passwd, tmp_pass)) { return true; }
	} return false;
}
bool signin(char *username, char *password)
{
	if(check(username, password))
	{
		printf("Hey %s, you logged in successfully\n", username);
		/* Uncomment the following line for double-free & use-after-free) */
		// free(username); free(password); 
		return true;
	} else { printf("The username or password is wrong\n"); return false; }
}
void authentication(char *username, char *password)
{
	/* Uncomment the following two lines for heap-based buffer overflow & use-after-free & double-free */
	// char* tmp_user = (char *)(malloc(80*sizeof(char)));
	// char* tmp_pass = (char *)(malloc(80*sizeof(char)));
	/* Uncomment the following line for stack-based buffer overflow */
	// char tmp_user[32], tmp_pass[32];
	/* POTENTIAL FLAW: data may not have enough space to hold source */
	/* Uncomment the following line for heap-based & stack-based buffer overflow */
	// memcpy(tmp_user, username, strlen(username));
	/* POTENTIAL FLAW: data may not have enough space to hold source */
	/* Uncomment the following line for heap-based & stack-based buffer overflow */
	// memcpy(tmp_pass, password, strlen(password));
	signin(tmp_user, tmp_pass);
	/* POTENTIAL FLAW: Use of data that may have been freed in line 63 */
	/* Uncomment the following line for use-after-free */
	// tmp_user[80-1] = '\0'; tmp_pass[80-1] = '\0';
	/* POTENTIAL FLAW: Free data here - line 63 frees data as well */
	/* Uncomment the following line for double-free */
	// free(tmp_user);	free(tmp_pass);
}
int main (int argc, char *argv[])
{
	/* uncomment for stack-based buffer overflow */
	// char username[64]; char password[64];
	/* Uncomment the following two lines for heap-based buffer overflow & use-after-free & double-free */
	// char *username = (char *)(malloc(100*(sizeof(char))));
	// char *password = (char *)(malloc(100*(sizeof(char))));
	printf("Enter username :"); scanf("%s", username); printf("Enter password :"); scanf("%s", password);
	if(argc >= 3) { authentication(argv[1], argv[2]); }
	else
	{
		if(username[0] >= 'A' && username[0] <= 'Z') { signup(username, password); }
		else { printf("The selected username is not valid, it must start with an uppercase letter"); }
	}
}
\end{lstlisting}
%\vspace{3mm}
\caption{Source codes of the four designed complex programs\label{fig:appendix}}
\end{figure*}
%\vfill\pagebreak

\bibliographystyle{splncs04}
\bibliography{Paper}
\clearpage

\end{document}